\documentclass[preprint,aps,prb]{revtex4-1}
\usepackage{graphics}
\usepackage{amssymb}
\usepackage{upgreek}
\usepackage[tight,TABTOPCAP]{subfigure}
\usepackage{subfloat}
\usepackage{float}
\usepackage{graphicx}
\usepackage{dcolumn}
\usepackage{bm}
\usepackage{setspace}
\usepackage{wrapfig}
\usepackage[intlimits]{empheq}
\usepackage{color}
\usepackage{relsize}

\makeatletter

\begin{document}

\title{Determination of boson spectrum from optical data in pseudogap phase of underdoped cuprates}
\author{Jungseek Hwang$^{1}$}
\email{jungseek@skku.edu; Corresponding author}
\author{J. P. Carbotte$^{2,3}$}

\affiliation{$^1$Department of Physics, Sungkyunkwan University, Suwon, Gyeonggi-do
440-746, Republic of Korea}
\affiliation{$^2$Department of Physics and Astronomy, McMaster University, Hamilton, Ontario L8S 4M1 Canada}
\affiliation{$^3$The Canadian Institute for Advanced Research, Toronto, ON M5G 1Z8 Canada}

\date{\today}
\begin{abstract}
Information on the nature of the dominant inelastic processes operative in correlated metallic systems can be obtained from an analysis of their AC optical response. An electron-boson spectral density can usefully be extracted. This density is closely related to the optical scattering rate. However, in the underdoped region of the high T$_c$ cuprate phase diagram a new energy scale (the pseudogap) emerges, which alters the optical scattering and needs to be taken into account in any fit to data. This can influence the shape and strength of the recovered boson spectral function. Including a pseudogap in an extended maximum entropy inversion for optimally doped Bi-2212 is more consistent with existing data than when it is left out as done previously.

\end{abstract}
%
\maketitle

\section{Introduction}

Boson structures have been noted in the physical properties of the high critical temperature superconducting cuprates using various techniques.\cite{carbotte:2011} These include angular resolved photoemission (ARPES), infrared optical conductivity (IR), point contact, as well as scanning tunneling spectroscopy (STM), and Raman. Assuming that these structures can be described approximately within a boson exchange formalism, they can provide valuable information on the effective underlying electron-boson spectral density $I^2\chi(\omega)$ related to the inelastic scattering or glue involved in their superconductivity. In such an approach a Kubo formula relates the spectral density of interest to the AC optical conductivity. For example, inversion of optical data then proceeds directly from the conductivity $\sigma(\omega)$ or from the optical scattering rate $1/\tau^{op}(\omega)$. A least square fit can be used to determine parameters in an assumed mathematical form for $I^2\chi(\omega)$. A maximum entropy technique based on simplified yet quite accurate analytic forms for the conductivity derived by Allen\cite{allen:1971} have also been employed. Such a technique has the advantage that there is no need for any assumption about the particular form for $I^2\chi(\omega)$. Instead it is obtained numerically in which case more fine details may emerge.

Much of the work so far has been restricted\cite{schachinger:2006,heumen:2009,hwang:2007} to cases in which the electronic structure does not develop a new energy scale of the same order of magnitude as the boson energies we wish to probe. In principle, the assumption that the electronic density of state is energy independent in the energy range of interest, is likely to be valid only for the optimal and overdoped region of the cuprate phase diagram. Modification can be expected in the underdoped region when a pseudogap develops\cite{hwang:2008a,hwang:2008,hwang:2008b}. Some analyses of data including a pseudogap have already appeared\cite{hwang:2006,hwang:2011}, in which parameters characterizing an assumed form for the pseudogap density of states as well as for the spectral density are varied in a least square fit. Here we consider how the maximum entropy technique\cite{schachinger:2006} is to be adapted to the case of an energy dependent density of states.

This work will be based on a generalized approximate, but still accurate, analytic form for the relationship between the optical scattering rate and the spectral density which further includes an electronic density of states factor $\tilde{N}(\omega)$.\cite{sharapov:2005} When this factor is assumed constant we recover the equation given by Allen.\cite{allen:1971} The new equation at zero temperature $T = 0$ was given by Mitrovic and Fiorucci\cite{mitrovic:1985} and later generalized by Sharapov and Carbotte\cite{sharapov:2005} to include finite temperatures. In the case of finite $T$ but constant $\tilde{N}(\omega)$. The generalized formula also reduces, as it should, to that given by Shulga et al.\cite{shulga:1991} as the finite temperature generalization of the original Allen equation.\cite{allen:1971}

Section II is an introduction to the theoretical concepts on which this work is based, and section III is a summary of the maximum entropy inversion method (MEM) used here. Section IV establishes preliminary numerical MEM results which will guide us in inversion of real data which is found in section V.

\section{Theoretical considerations}

The frequency dependent optical conductivity $\sigma(T,\omega)$ for a correlated electron system can usefully be written in terms of a frequency and temperature dependent optical self energy $\Sigma^{op}(T,\omega)$ which plays a role in optics similar to the quasiparticle self energy of angular resolved photoemission (ARPES). Denoting the plasma energy by $\Omega_p$ we can write
\begin{equation}\label{eqn1}
\sigma(T,\omega)=\frac{i}{4 \pi}\frac{\Omega_p^2}{\omega-2\Sigma^{op}(T,\omega)}
\end{equation}
The imaginary part of $-2\Sigma^{op}(T,\omega)$ defines an optical scattering rate $1/\tau^{op}(T,\omega)$ and the real part a renormalized optical effective mass $m^{*op}(T,\omega)/m$ with $\omega[m^{*op}(T,\omega)/m -1]=-2Re\Sigma^{op}(T,\omega)$. The optical mass enhancement $\lambda^{op}(T,\omega)$ is defined as $1+\lambda^{op}(T,\omega)=m^{*op}(T,\omega)/m$. In terms of $1/\tau^{op}(T,\omega)$ and $\lambda^{op}(T,\omega)$ the conductivity takes on a Drude form with its real part
\begin{equation}\label{eqn2}
\sigma_1(T,\omega)=\frac{\Omega_p^2}{4 \pi} \frac{1/\tau^{op}(T,\omega)}{[\omega(1+\lambda^{op}(T,\omega))]^2+[1/\tau^{op}(T,\omega)]^2}
\end{equation}
which differs from its simplest rendition only through energy and temperature dependence\cite{carbotte:1995,nicol:1991,nicol:1991:2,schachinger:1997} in $1/\tau^{op}(T,\omega)$ and mass renormalization $\lambda^{op}(T,\omega)$. This energy and temperature dependence carries the information on the inelastic scattering here, due to coupling to an effective boson exchange mechanism. In conventional superconductors these lead to so called strong coupling corrections\cite{carbotte:1986,marsiglio:1992,mitrovic:1980,schachinger:1990} to conventional BCS theory. Of course additional corrections can also play a role such as energy dependence\cite{schachinger:1990,arberg:1993,mitrovic:1983,mitrovic:1983:2} in the density of electronic states and momentum anisotropies.\cite{branch:1995,leung:1976,odonovan:1995} In general $\sigma_1(T,\omega)$ of eqn.(\ref{eqn2}) can be calculated from a Kubo formula \cite{carbotte:1995,nicol:1991,nicol:1991:2,schachinger:1997} which involves, in a bubble approximation, thermal factors and the product of two single particle spectral functions $A(k,\omega)$ both at the same momentum $k$ but displaced in energy $\omega$ by the photon energy $\Omega$, neglecting vertex corrections. For a boson exchange model with electron-boson spectral density $I^2\chi(\omega)$, Allen\cite{allen:1971} derived a very simple approximate, but analytic, formula which relates $1/\tau^{op}(T,\omega)$ directly to $I^2\chi(T,\omega)$. It was generalized to finite temperature by Shulga et al.\cite{shulga:1991} and by Sharapov and Carbotte\cite{sharapov:2005} to the case when there is important energy dependence in the effective electronic density of state $\tilde{N}(\omega)$ which needs to be taken into account. The formula of Sharapov and Carbotte\cite{sharapov:2005} is
\begin{eqnarray}\label{eqn3}
1/\tau^{op}&=&\frac{\pi}{\omega}\int^{\infty}_{0} d\Omega I^2\chi(\Omega)\int^{+\infty}_{-\infty} dz [N(z-\Omega)+N(-z+\Omega)] \nonumber \\
&\times& [n_B(\Omega)+1-f(z-\Omega)][f(z-\omega)-f(z+\omega)]
\end{eqnarray}
where $n_B(\Omega)$ and $f(\Omega)$ are respectively the Bose-Einstein and Fermi-Dirac distribution functions. This generalized formula properly reduces to the form given by Shulga el al.\cite{shulga:1991} when the effective density of state $\tilde{N}(z)$ is constant, and also to Allen's form when temperature is taken to be zero. For zero temperature but a variable density of state, we get the formula given by Mitrovic and Fiorucci\cite{mitrovic:1985}
\begin{equation}\label{eqn4}
\frac{1}{\tau^{op}(T=0,\omega)}\equiv\frac{1}{\tau^{op}(\omega)}=\frac{2 \pi}{\omega}\int^{\omega}_{0} d\Omega I^2\chi(\Omega)\int^{\omega-\Omega}_{0}d z\tilde{N}(z)
\end{equation}
where $\tilde{N}(\omega)$ is the symmetrized density of state $[N(\omega)+N(-\omega)]/2$. When it is constant and equal to one we obtain the well known Allen formula and find that the second derivative\cite{marsiglio:1998} of $\omega/\tau^{op}(\omega)$ is $I^2\chi(\omega)$ i.e.
\begin{equation}\label{eqn6}
\frac{1}{2\pi}\frac{d^2}{d\omega^2}\Big{[}\frac{\omega}{\tau^{op}(\omega)}\Big{]}=I^2\chi(\omega)
\end{equation}
While formula (\ref{eqn6}) is simple, even when the full Kubo formula for the conductivity in a boson exchange model is used, this formula is known to reproduce accurately the spectral density in the energy range in which it is non-zero. Above the cutoff in $I^2\chi(\omega)$ the derivative on the left hand side of eqn.(\ref{eqn6}) can become negative in the more complete theory\cite{marsiglio:1998}, but this is of no consequence here.

\section{Maximum entropy inversion with energy dependent electronic density of state}

Equation (\ref{eqn4}) can be written in the general form
\begin{equation}\label{eqn7}
\frac{1}{\tau^{op}(\omega)} =\int d\Omega I^2\chi(\Omega) K(\omega,\Omega)
\end{equation}
where the kernel $K(\omega,\Omega)$ can be read off eqn.(\ref{eqn4}) but for the present purpose can be left general and unspecified. For a general kernel, $K(\omega,\Omega)$, and input data, $D_{in}(\omega)$, with $D_{in}(\omega)=\int^{+\infty}_{0} K(\omega,\Omega)I^2\chi(\Omega)d\Omega$ the deconvolution of this equation to recover an effective spectral density, $I^2\chi(\Omega)$ is ill-conditioned and here we use a maximum entropy technique.\cite{shulga:1991} The equation can be discretized $D_{in}(i)=\sum_{j}K(i,j)I^2\chi(j)\Delta\Omega$ where $\Delta\Omega$ is the differential increment on the integration over $\Omega_j=j\Delta\Omega$ with $j$ an integer. We define a $\chi^2$ by
\begin{equation}\label{eqn8}
\chi^2=\sum^{N}_{i=1}\frac{[D_{in}(i)-\Sigma(i)]^2}{\epsilon_i^2}
\end{equation}
where $D_{in}(i)$ is the input data, and $\Sigma(i)\equiv\sum_{j}K(i,j)I^2\chi(j)\Delta\Omega$ is calculated from the known kernel and a given choice of $I^2\chi(j)$, and $\epsilon_i$ is the error assigned to the data $D_{in}(i)$. Constraints such as positive definiteness for the boson exchange function are noted and the entropy functional
\begin{equation}\label{eqn9}
L=\frac{\chi^2}{2}-\sigma S
\end{equation}
is minimized with the Shannon-Jones entropy\cite{schachinger:2006}, $S$
\begin{equation}\label{eqn10}
S=\int^{\infty}_{0}\Big{[}I^2\chi(\Omega)-m(\Omega)-I^2\chi(\Omega)\ln\Big{|}\frac{I^2\chi(\Omega)}{m(\Omega)}\Big{|} \Big{]}d\Omega
\end{equation}
The parameter $\sigma$ in eqn.(\ref{eqn9}) controls how close a fit to the data is obtained. The parameter $m(\Omega)$ is here taken to be some constant value on the assumption that there is no a priori knowledge of the functional form of the electron-boson spectral density $I^2\chi(\Omega)$. While there is no guarantee that a boson exchange model can successfully reproduce consistently, quantitatively, and accurately all the details of optical data in highly correlated systems, it does produce important information. An important fact to note, and this has been well documented and stressed in the review of Carbotte, Timusk, and Hwang\cite{carbotte:2011}, is that there is a great deal of qualitative agreement between recovered spectrum using ARPES, IR, Raman and STM tunneling. This provides confidence to go further and now include more rigorously pseudogap features which here enter in eqn.(\ref{eqn7}) through the density of state factor $\tilde{N}(\omega)$ with the necessary modifications due to the opening of a pseudogap $\Delta_{pg}$. Here we do not wish to commit to a particular specific pseudogap model but instead take a parameterized form for the effective symmetrized DOS $\tilde{N}(\omega)$ and vary parameters. Once this is fixed, maximum entropy will provide us with a spectral density $I^2\chi(\omega)$ for a given set of data for the optical scattering rate $1/\tau^{op}(\omega)$. This does not tell us anything about the actual origin of the boson involved in the scattering of the charge carriers, and the origin of these bosons remains controversial. A review of all available inversions based on optics as well as on Raman and angular resolved photo emission (ARPES) and other considerations given in reference [1] led the authors to nevertheless conclude that the spin fluctuations play the major role with possibly a small\cite{schachinger:2010} contribution at the 10 \% level from the phonons. Should the recently\cite{Li:2010,Almeida:2012,Li:2012} discovered novel magnetic modes associated with zero momentum ($q = 0$) contribute significantly to the glue, they would also in principle, be included in the recovered spectra.

\section{Preliminary numerical results for maximum entropy inversions}

Taking the second derivative of $\omega/\tau^{op}(\omega)$ in eqn.(\ref{eqn4}) gives
\begin{equation}\label{eqn14}
\frac{1}{2\pi}\frac{d^2}{d\omega^2}\Big{[}\frac{\omega}{\tau^{op}(\omega)}\Big{]}= I^2\chi(\omega)\tilde{N}(0)-\int^{\omega}_{0}d\Omega I^2\chi(\Omega)\frac{d}{d\Omega}[\tilde{N}(\omega-\Omega)]
\end{equation}
which is quite different from the result of eqn.(\ref{eqn6}) for the constant density of state case. Here the first term does give $I^2\chi(\omega)$ reduced by the factor $\tilde{N}(0) \equiv N_0$ and the second is a correction. It is instructive to change the integral in eqn.(\ref{eqn14}) through an integration by parts to obtain
\begin{equation}\label{eqn15}
\frac{1}{2\pi}\frac{d^2}{d\omega^2}\Big{[}\frac{\omega}{\tau^{op}(\omega)}\Big{]}= I^2\chi(\omega=0)\tilde{N}(\omega)+ \int^{\omega}_{0} d\Omega \tilde{N}(\omega-\Omega) \frac{d}{d\Omega}[I^2\chi(\Omega)]
\end{equation}
This form provides a first term for the second derivative of $\omega/\tau^{op}(\omega)$ which is now the product of $I^2\chi(\omega)$ at $\omega=0$ and $\tilde{N}(\omega)$ while the second term is a correction to this simplified result. It is interesting to consider the case of a marginal fermi liquid (MFL) model for the spectral density $I^2\chi(\Omega)$ which has the form $A\tanh(\Omega/k_B T)$. For low temperature this form provides a constant $I^2\chi(\omega) = A$ and its derivative is zero. Thus for this particular case the second derivative of eqn.(\ref{eqn15}) gives the product of $I^2\chi(\omega=0)\tilde{N}(\omega)=I^2\chi(\omega)\tilde{N}(\omega)$ and is to be contrasted with the eqn.(\ref{eqn6}). For a constant $\tilde{N}(\omega)$ we get $I^2\chi(\omega)$ while for a constant spectral density we get the effective density of states $\tilde{N}(\omega)$ which includes the pseudogap. It is important to stress that this result is restricted to a constant spectral density and the second term in eqn.(\ref{eqn15}) will provide modifications in all other cases.

\begin{figure}
\vspace*{-1.4cm}%
\centerline{\includegraphics[width=4.5in]{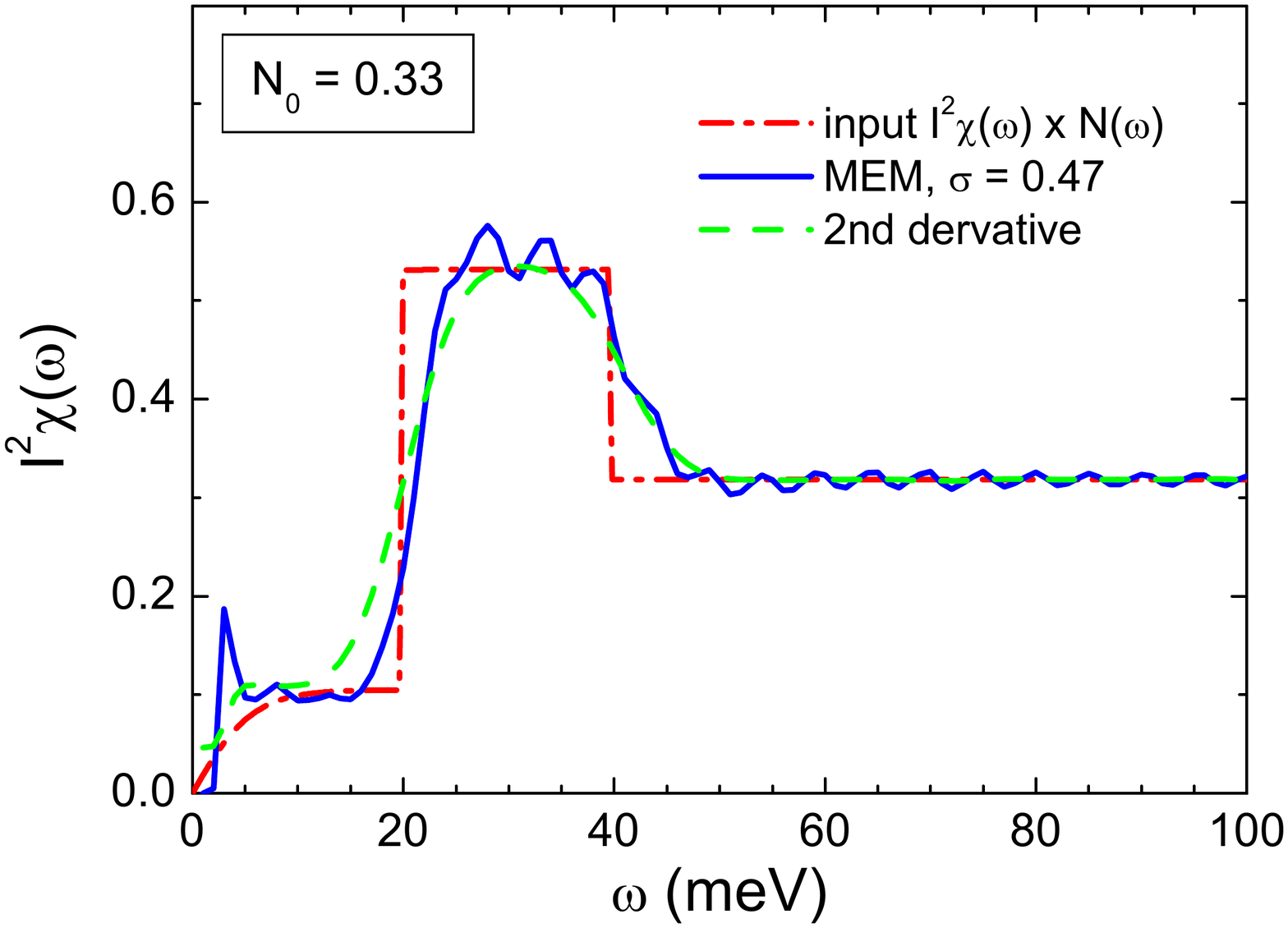}}
\vspace*{-1.0cm}%
\caption{(Color online) The product (dash dotted red curve) of the input marginal fermi liquid (MFL) spectral density $I^2\chi(\omega)$ multiplied by a density of state $\tilde{N}(\omega)$ which is 33 \% its constant energy value below $\omega=\Delta_{pg} = 20$ meV. Lost states in $\tilde{N}(\omega)$ are piled up above $\Delta_{pg}$ between 20 and 40 meV. The solid blue curve is the spectrum recovered from a maximum entropy inversion of the scattering rate $1/\tau^{op}(\omega)$ and the dashed green curve is the second derivative of eqn.(\ref{eqn14}) and (\ref{eqn15}).}
\label{fig1}
\end{figure}

In Fig. \ref{fig1} we show results of our maximum entropy inversions of optical data generated in a MFL model for $I^2\chi(\omega)$, and square well model for $\tilde{N}(\omega)$ which is taken equal to 0.33 below the pseudogap energy $\omega=\Delta_{pg}= 20$ meV with missing states piled up just above this energy and distributed equally in the range $\omega=\Delta_{pg}$ to $2\Delta_{pg}$. The input product of $I^2\chi(\omega)\times \tilde{N}(\omega)$ is represented by the red dash-dotted curve, the maximum entropy inversion is the solid blue curve, and the second derivative technique yields the dashed green curve. Both agree quite well with the input product; our expectation that we should get to a good approximation to $I^2\chi(\omega)\times \tilde{N}(\omega)$ is borne out by the numerical work. It is clear however that we cannot get independent information on $\tilde{N}(\omega)$ and $I^2\chi(\omega)$ from optics even in this very simplified case.

\section{Application of maximum entropy method to real data}

\begin{figure}
\vspace*{-1.4cm}%
\centerline{\includegraphics[width=4.0in]{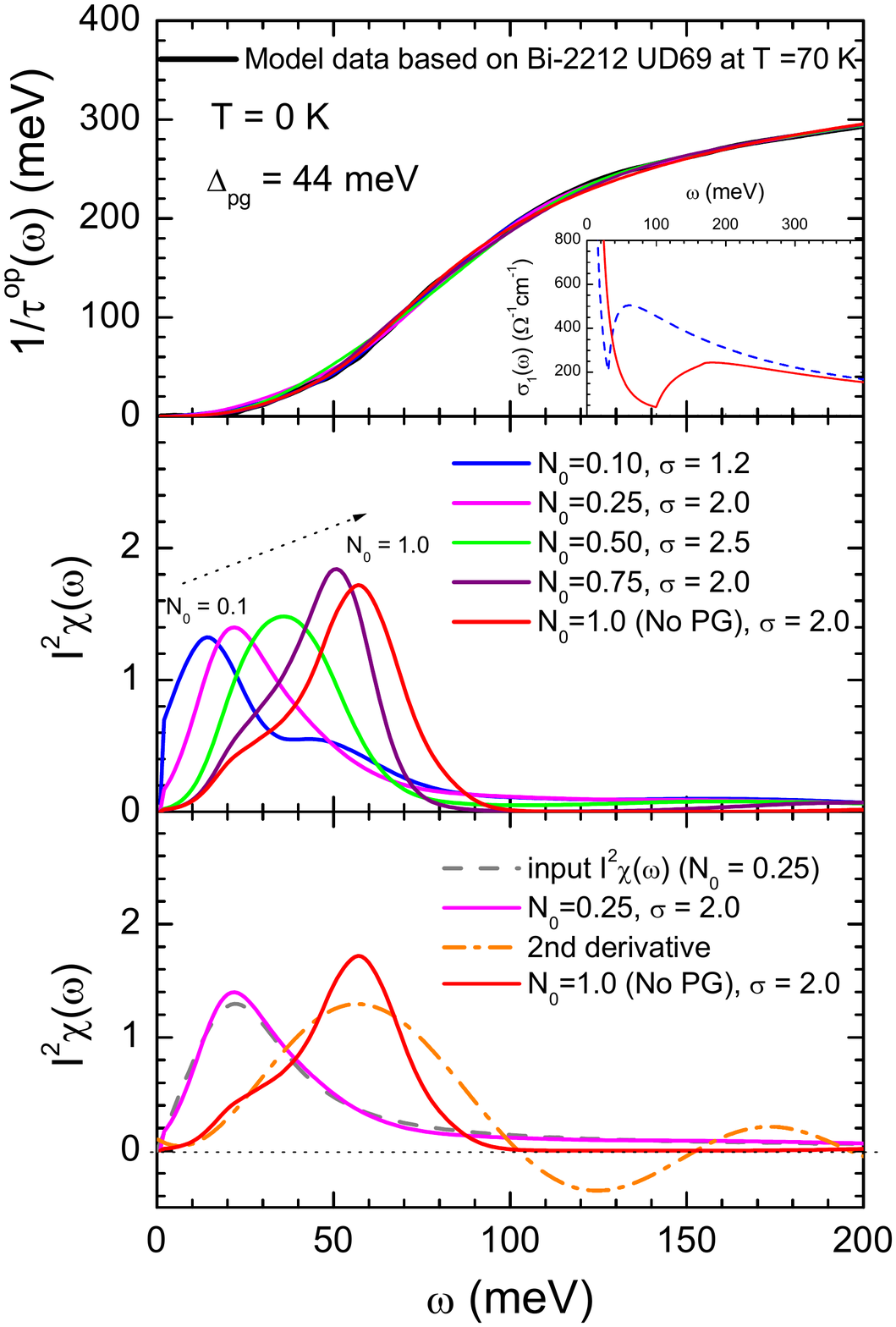}}
\vspace*{-1.0cm}%
\caption{(Color online) Model optical scattering rate data $1/\tau^{op}(\omega)$ (solid black curve) for zero temperature based on a Bi-2212 UD69 sample (top frame). The other curves are our maximum entropy (ME) fits. The middle frame gives the recovered electron-boson spectral density $I^2\chi(\omega)$ when our ME inversion includes a pseudogap $\Delta_{pg} = 44$ meV with various values of $N_0$ as noted in the figure. The bottom frame is for fix $N_0 = 0.25$. The grey dashed curve gives our input model for $I^2\chi(\omega)$ and the solid purple curve the spectrum recovered from ME inversion including the model DOS $\tilde{N}(\omega)$ with pseudogap. The solid red curve is the spectrum recovered when $\tilde{N}(\omega)$ is assumed constant i.e. $N_0 = 1.0$ in the inversion process and the dash-dotted orange curve is the second derivative result $1/(2\pi) d^2/d\omega^2[\omega/\tau^{op}(\omega)]$. In the inset we display the real part of the optical conductivity $\sigma_1(\omega)$ with ($N_0 =$ 0.0, solid red) and without ($N_0 =$ 1.0, dashed blue) pseudogap from the work in reference [6].}
\label{fig2}
\end{figure}

\begin{figure}
\vspace*{-1.4cm}%
\centerline{\includegraphics[width=4.0in]{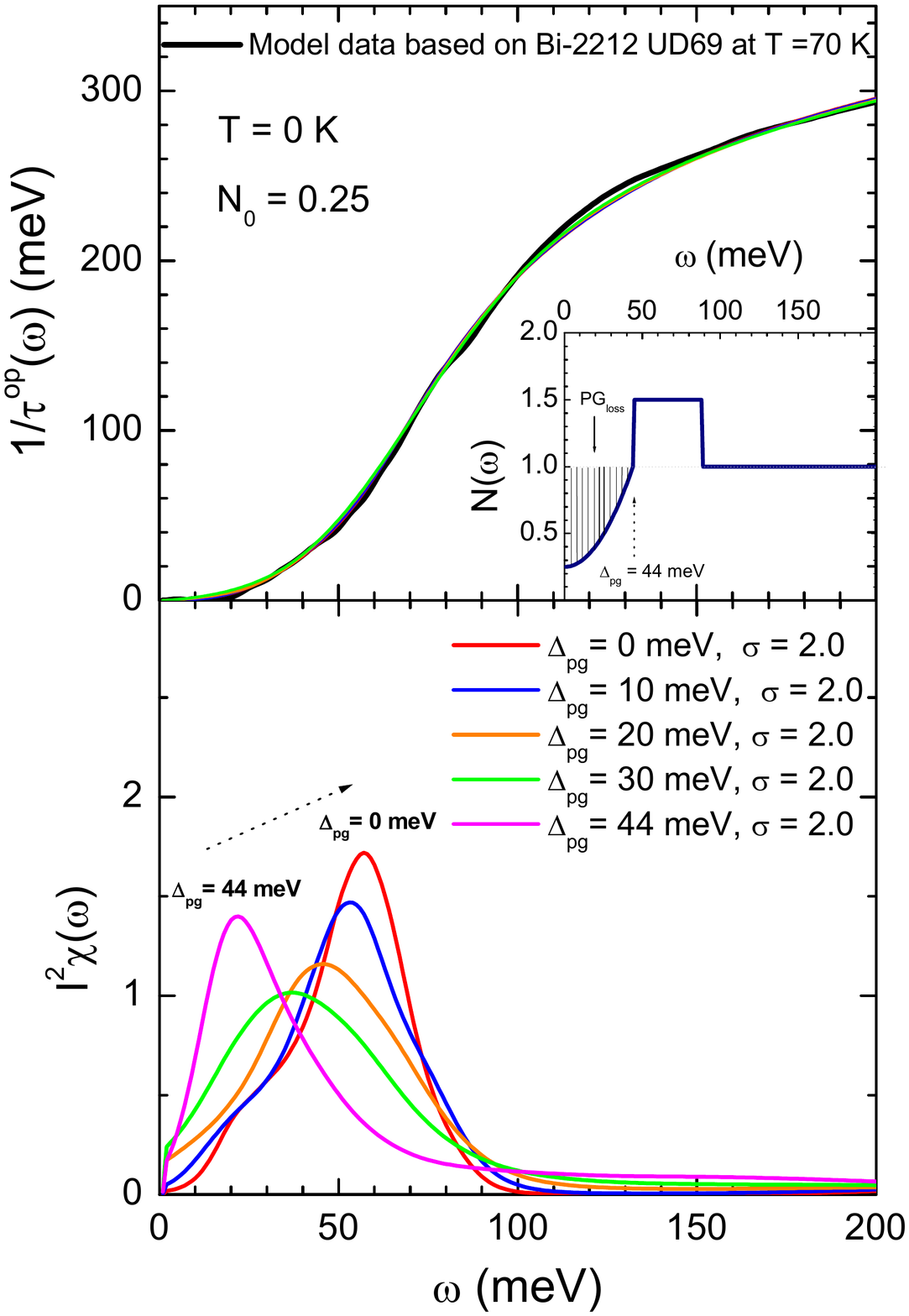}}
\vspace*{-1.0cm}%
\caption{(Color online) As in Fig. \ref{fig2} but now the depth of the well in the density of state $\tilde{N}(\omega)$ is kept fixed at $N_0 = 0.25$ and the size of the pseudogap $\Delta_{pg}$ is varied as indicated in the figure. The various lines in the top frame for the optical scattering rate $1/\tau^{op}(\omega)$ are the fits to the input data (black curve). The inset shows the model density of state used for the pseudogap and the shaded region defines the spectral weight lost below $\Delta_{pg}$ due to pseudogap formation which we denote by $PG_{loss}$.}
\label{fig3}
\end{figure}

In the top frame of Fig. \ref{fig2} we present a model\cite{hwang:2011} for the optical scattering rate $1/\tau^{op}(\omega)$ serving as a convenient input for our maximum entropy inversions. The model is based on data for Bi-2212 UD 69 at $T =$ 70 K\cite{hwang:2007,hwang:2011} with intercept at $\omega=0$ set zero so as to simulate a clean sample at zero temperature. Based on this realistic form we now study how maximum entropy inversion at $T =$ 0 works when there is a pseudogap in the system but the corresponding details of the density of states $\tilde{N}(\omega)$ are not known. In the middle frame we show the recovered $I^2\chi(\omega)$ for 5 cases. In all instances we have taken a pseudogap model previously used with success by Hwang\cite{hwang:2011} in his least square fit analysis of the Bi-2212. The model has the from.\cite{hwang:2006,hwang:2011}
\begin{eqnarray}\label{eqn16}
\tilde{N}(\omega) &=& N_0+(1-N_0)\Big{(}\frac{\omega}{\Delta_{pg}}\Big{)}^2 \:\:\:\:\:   \mbox{for} \:\: |\omega|\leq \Delta_{pg} \nonumber \\ &=& 1+\frac{2(1-N_0)}{3} \:\:\:\:\:   \mbox{for} \:\: |\omega|\in(\Delta_{pg}, 2\Delta_{pg}) \nonumber \\ &=& 1 \:\:\:\:\:   \mbox{for} \:\: |\omega|\geq 2\Delta_{pg}.
\end{eqnarray}
This mathematical form is illustrated in the inset of Fig. \ref{fig3} for a case $N_0 = 0.25$ and $\Delta_{pg} = $44 meV. In Hwang's previous work\cite{hwang:2011} the electron-boson spectral density was modeled with two analytic curves
\begin{equation}\label{eqn17}
I^2\chi(\omega)=\frac{A_s \omega}{\omega_s^4+\omega^4}+\frac{A_m \omega}{\omega_m^2+\omega^2}
\end{equation}
which consists of an MMP piece\cite{millis:1990} (second term) representing coupling to spin fluctuations as in the work of Millis, Monien and Pines (MMP). This provides a background extending over several 100 meV with $\omega_m$ a spin fluctuation frequency and $A_m$ an amplitude. An additional sharp peak (first term), possibly representing coupling to an optical resonance at $\omega_s$, is also included in the least square fit to the scattering rate which has six parameters. Here, however, we will not use the functional form eqn.(\ref{eqn17}) for $I^2\chi(\omega)$ but instead employ a maximum entropy technique; this in no way commits us to a particular form for $I^2\chi(\omega)$. Such a technique applied to optical data in La$_{1.83}$Sr$_{0.17}$CuO$_4$ produced a two peak structure in the recovered electron-boson spectral density for example\cite{hwang:2008c}.  Most recovered spectra\cite{hwang:2007,yang:2009,heumen:2009}, however, show a low energy resonance structure superimposed on a background which extends to energies as high as 300 meV or even higher. Such features are well represented qualitatively with the analytic form of eqn.(\ref{eqn17}). We note that the density of state model used in reference \cite{hwang:2011}, which we retain here, is similar to what is found in STM work of Renner et al.\cite{renner:1998}  The $I^2\chi(\omega)$ obtained by Hwang\cite{hwang:2011} is shown as the grey dashed line in the bottom frame of Fig. \ref{fig2} and will be discussed later. In the middle frame we show results of maximum entropy inversion of eqn.(\ref{eqn7}) with kernel given in eqn.(\ref{eqn4}) and the model $\tilde{N}(\omega)$ as in Hwang\cite{hwang:2011} and previously in Hwang et al.\cite{hwang:2008b} where it is applied to the analysis of OrthoII YBCO. In all cases $\Delta_{pg} = 44$ meV, but various value of $N_0$ in eqn.(\ref{eqn16}) are employed, namely blue ($N_0$ = 0.1), pink ($N_0 = 0.25$), green ($N_0$ = 0.5), black ($N_0$ = 0.75), and red $N_0$ = 1.0 which corresponds to the case of no pseudogap i.e. a flat density of state. In all instances good fit to $1/\tau^{op}(\omega)$ is obtained as shown in the top frame. We see that the recovered $I^2\chi(\omega)$ however changes significantly as $N_0$ goes from 0.1 to 1.0. The peak moves to higher frequency and generally increases in height and more spectral weight is transferred from the $\omega \cong 0$ region with increasing $N_0$. If there were a pseudogap in the system with $N_0 = 0.25$ as found in the least square fit approach of Hwang\cite{hwang:2011}, and it were ignored in a maximum entropy fit, we see that the resultant $I^2\chi(\omega)$ given in the dashed grey curve of the lower frame would be very different from its true value. It is clear from these results that, in an analysis of optical data in the underdoped region of cuprate phase diagram, we need to include the pseudogap if we are to obtain a reliable value of the spectral density and in particular get correctly the position of its peak which represents coupling to a sharp resonance mode. The inset of the top frame makes a similar point. It shows the real part of the optical conductivity based on a model $I^2\chi(\omega)$ of reference [6] (solid red) including a pseudogap and without (dashed blue). We see significant differences between these two curves. In particular the effective boson assisted incoherent Holstein sideband shows a sharp onset at the energy of the peak in our model $I^2\chi(\omega)$, with onset is shifted to the right by the pseudogap energy as compared with the case without pseudogap. It is also reduced in magnitude.

In the bottom frame of Fig. \ref{fig2} we show that when maximum entropy is used for inversion with the known pseudogap model, we get an excellent reproduction (solid purple curve) of its least square fit determination (dashed grey curve). On the other hand if the maximum entropy inversion proceeds on the assumption of a constant density of state we get the solid red curve which peaks at higher energy than does the input $I^2\chi(\omega)$. This agrees well with the second derivative result shown as the orange dash-dotted curve. Fig. \ref{fig3} provides additional results. The lower frame gives our MEM results for $I^2\chi(\omega)$ when various values of $\Delta_{pg}$ itself are used (pink 0 meV, blue 10 meV, orange 20 meV, blue 30 meV and red 44 meV, as before). The fixed parameter is the depth of the pseudogap well at $\omega = 0$ i.e. $N_0 = 0.25$ in all cases. The fits to the scattering rate data are given in the top frame. What is clear from these data is that decreasing the value of the pseudogap pushes the peak in the MEM $I^2\chi(\omega)$ to higher energies, as the spectral density tries to compensate for the loss in scattering implied by a decrease in $\Delta_{pg}$.

\begin{figure}
\vspace*{-1.4cm}%
\centerline{\includegraphics[width=4.0in]{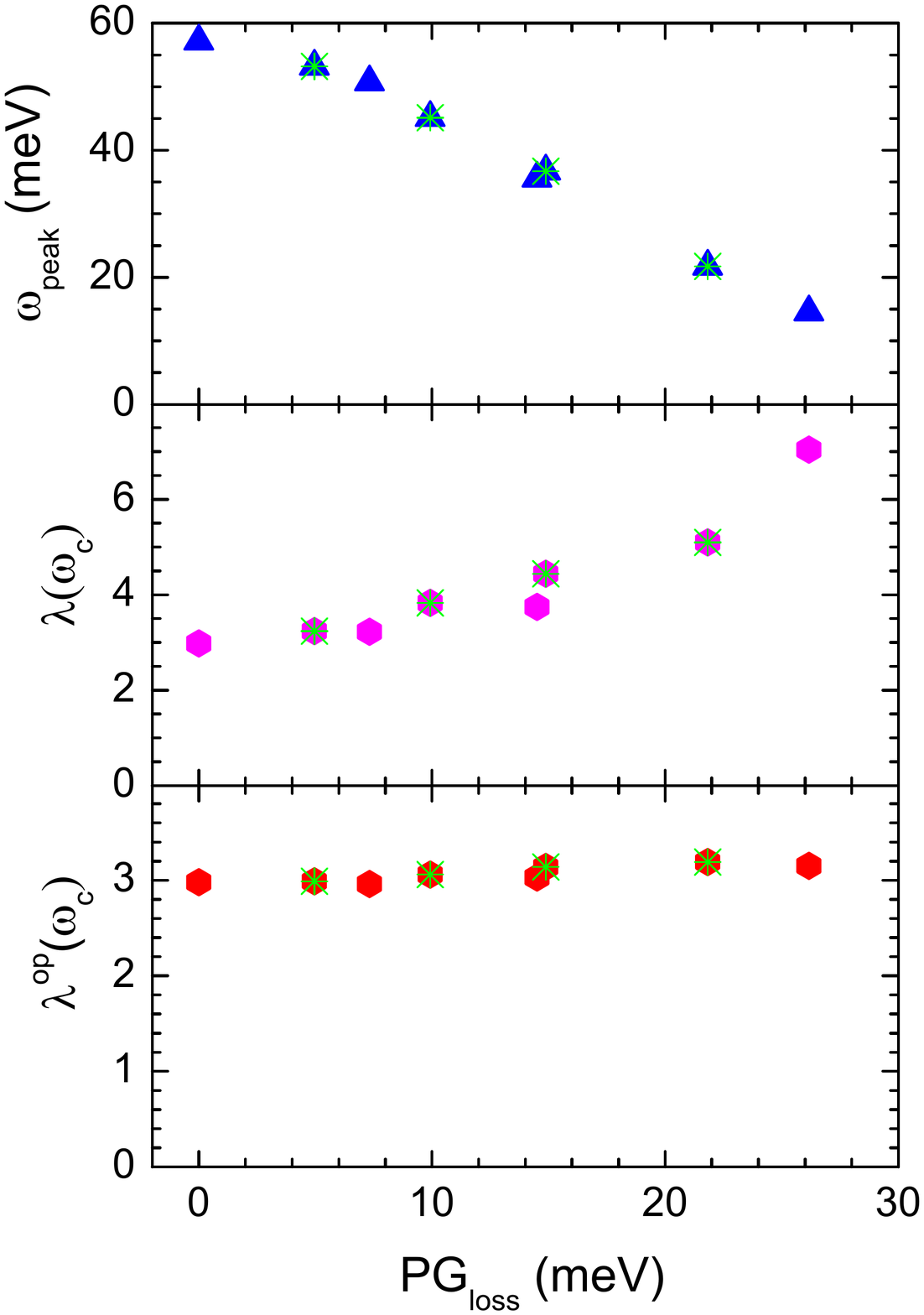}}
\vspace*{-1.0cm}%
\caption{(Color online) Microscopic parameters associated with our recovered $I^2\chi(\omega)$ spectra as a function of $PG_{loss}$ in meV. The top frame gives the energy of the peak in the spectral density. The middle frame gives the spectral mass enhancement $\lambda(\omega_c)$ defined as twice the first inverse moment of $I^2\chi(\Omega)$. The various points shown are based on the data in Fig. \ref{fig2} and Fig. \ref{fig3}. The bottom frame gives the optical mass enhancement factor $\lambda^{op}(\omega)$ at $\omega = 0$ of equations (\ref{eqn19}) and (\ref{eqn20}),  which is also the same as its quasiparticle renormalization. Both differ from $\lambda(\omega_c)$ when the electronic density of states varies with energy due to a finite pseudogap. }
\label{fig4}
\end{figure}

Plotting the position of the peak in $I^2\chi(\omega)$ obtained in all the cases considered in Fig. \ref{fig2} and Fig. \ref{fig3}, and additional ones for $N_0 = 0.25$ in the upper frame of Fig. \ref{fig4}, shows that they vary mainly with value of $PG_{loss}$ defined as the area of the shaded region in the density of state shown in the inset of Fig. \ref{fig3}. This represents the area lost in the density of state below the pseudogap energy $\omega = \Delta_{pg}$ as compared with its constant value (1.0 in our case). It is also the area recovered in our model above $\omega = \Delta_{pg}$ in the region to $2\Delta_{pg}$. The almost linear drop in the position of $\omega_{peak}$ as a function of increasing $PG_{loss}$ is a useful observation because it can be employed, as we will elaborate below, to constrain parameters in the effective density of state $\tilde{N}(\omega)$ when otherwise nothing is known about its variation with $\omega$. However, before we address this issue we show in the middle frame of Fig. \ref{fig4} corresponding results for the derived mass enhancement parameter $\lambda(\omega_c)$ defined in the usual way, as twice the first inverse moment of the spectral function i.e. $\lambda(\omega_c) = 2\int^{\omega_c}_{0} d\Omega \: I^2\chi(\Omega)/\Omega$ with a cutoff on $\Omega$ set to 5000 cm$^{-1}$. We will refer to this quantity as the spectral lambda. By its definition this is the electron-boson mass renormalization which enters many quantities such as the critical temperature and quasiparticle, and optical mass in the case of a flat density of electronic states. When $\tilde{N}(\omega)$ is not constant because there is a pseudogap, the optical and quasiparticle mass remain equal to each other, but are not given by the spectral lambda $\lambda(\omega_c)$. In our model for the optical conductivity $\lambda^{op}(\omega)$ is\cite{hwang:2008a,hwang:2008,hwang:2008b,knigavko:2005,knigavko:2006}
\begin{equation}\label{egn18}
\lambda^{op}(\omega)=\frac{2}{\omega^2}\int^{\omega_c}_{0} d\Omega I^2\chi(\Omega)\int^{\infty}_{0} d\omega'\tilde{N}(\omega')\ln\Big{[} \frac{(\omega'+\Omega)^2}{(\omega'+\Omega)^2-\omega^2}\Big{]}
\end{equation}
and its zero energy limit $\omega \rightarrow 0$ is
\begin{equation}\label{eqn19}
\lambda^{op}(\omega = 0) = 2\int^{\infty}_{0} d\omega' \tilde{N}(\omega') \int^{\omega_c}_{0} d\Omega\frac{I^2\chi(\Omega)}{(\omega'+\Omega)^2}
\end{equation}
which is different from the spectral lambda $\lambda(\omega_c)$ as discussed in reference\cite{hwang:2008a} and seen in the lower frame of Fig. \ref{fig4}. We can rewrite eqn.(\ref{eqn19}) for our $\tilde{N}(\omega)$ which is specified in eqn.(\ref{eqn16}).
\begin{eqnarray}\label{eqn20}
\lambda^{op}(\omega = 0) &=& 2\int^{\omega_c}_{0}d\Omega I^2\chi(\Omega) \Big{\{}N_0 \Big{(}\frac{1}{\Omega}-\frac{1}{\Omega+\Delta_{pg}}\Big{)} \nonumber \\
&+&(1-N_0)\Big{[} \frac{1}{\Delta_{pg}}+\frac{2\Omega}{\Delta_{pg}^2}\ln\Big{|}\frac{\Omega}{\Omega+\Delta_{pg}}\Big{|}
+\Big{(}\frac{\Omega}{\Delta_{pg}}\Big{)}^2\Big{(}\frac{1}{\Omega}-\frac{1}{\Omega+\Delta_{pg}}\Big{)}\Big{]}
\nonumber \\ &+&\Big{[}1+\frac{2}{3}(1-N_0)\Big{]}\Big{(}\frac{1}{\Omega+\Delta_{pg}}-\frac{1}{\Omega+2\Delta_{pg}}\Big{)}
+\frac{1}{\Omega+2\Delta_{pg}} \Big{\}}
\end{eqnarray}
We see that the spectral renormalization $\lambda(\omega_c)$increases with increasing $PG_{loss}$ (middle frame in Fig. \ref{fig4}), by contrast the optical mass is nearly independent of pseudogap details. As shown in the top frame, there is a decrease in $\omega_{peak}$ with increasing $PG_{loss}$ and this leads to an increased contribution to $\lambda(\omega_c)$ because of the term $1/\Omega$ in its definition. But in $\lambda^{op}(\omega=0)$ the additional presence of the pseudogap has the opposite tendency, because it reduces the effectiveness of small $\Omega$ below $\Delta_{pg}$ and both effects combined leave $\lambda^{op}(0)$ fairly constant.

\begin{figure}
\vspace*{-1.4cm}%
\centerline{\includegraphics[width=4.0in]{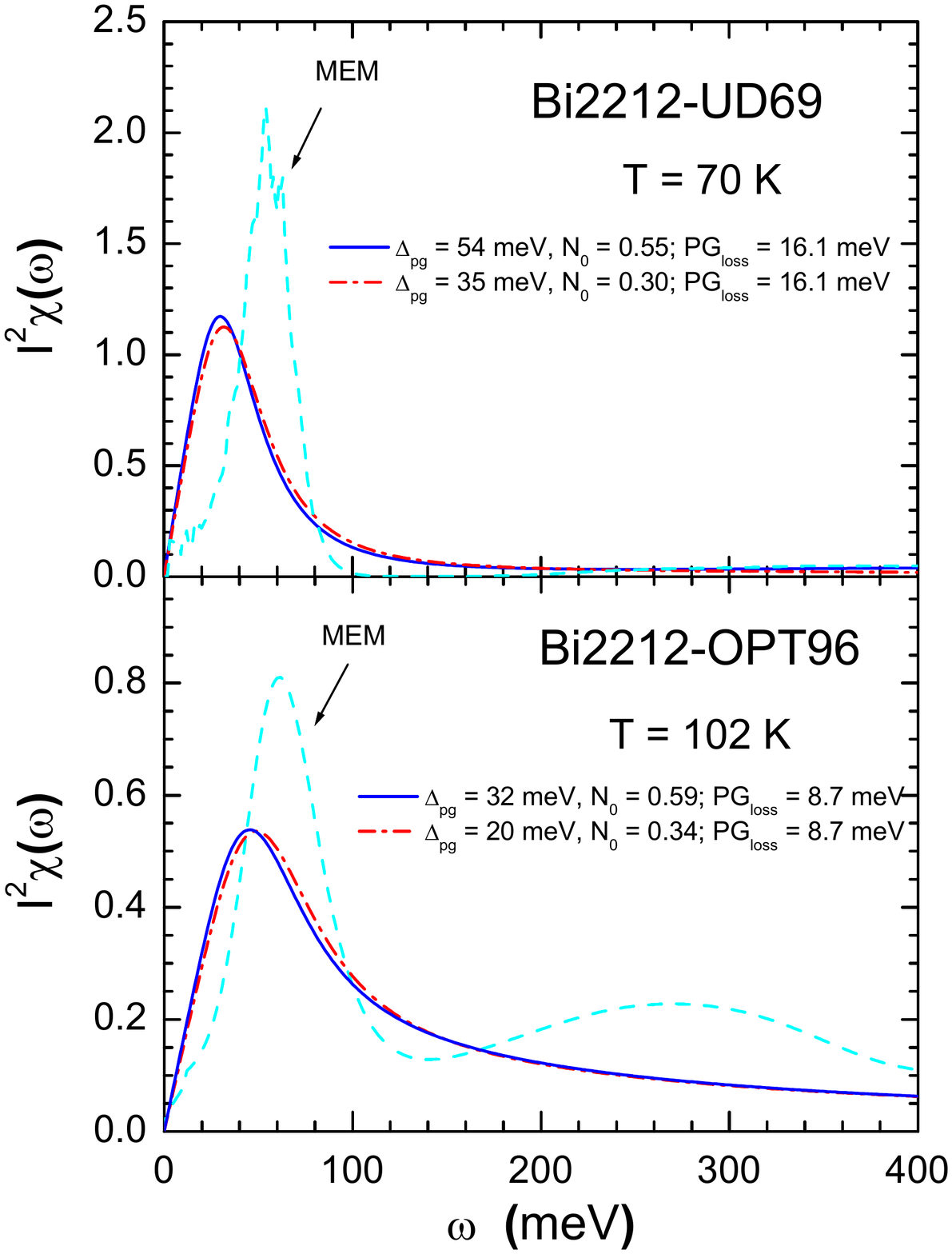}}
\vspace*{-1.0cm}%
\caption{(Color online) The electron-boson spectral density $I^2\chi(\omega)$ as function of energy $\omega$ in meV recovered from optical scattering rate data in Bi-2212. The upper frame is for an underdoped sample UD69 and the lower for optimally doped OPT96 from reference.\cite{hwang:2007} The dash-dotted red and solid blue curves result from a least square fit using a model spectral density consisting of an MMP background augmented with a sharp peak at $\omega_s/(3^{1/4})$ (see eqn. \ref{eqn17})\cite{hwang:2011}. This energy is taken to be 5.4 $k_B T_c$ and is the energy of the spin one resonance seen in the spin fluctuation spectrum by inelastic polarized neutron scattering. In addition a pseudogap is included in the density of state model shown in the inset of Fig. \ref{fig3} with parameter $N_0$ and $\Delta_{pg}$ which are also included in the least square fit with fixed value of $PG_{loss}$, 16.1 meV for UD69 and 8.7 meV for OPT96. The dashed blue curve is for comparison and is the spectrum obtained in a flat band maximum entropy inversion of the same data.}
\label{fig5}
\end{figure}

Armed with the observation that $\omega_{peak}$ decreases with $PG_{loss}$, and that this relationship is robust and minimally dependent on the details of the energy variation assumed for $\tilde{N}(\omega)$, we turn to experiments. In the upper frame of Fig. \ref{fig5} we reconsider the Bi-2212 UD69 first analyzed by Hwang\cite{hwang:2011}. Here we assume that the sharp peak in $I^2\chi(\omega)$ is due to coupling of the charge carriers to the spin one resonance observed in inelastic spin polarized neutron scattering\cite{he:2001,he:2002} following the law $\omega_{sr} \cong 5.4 k_B T_c$ where $T_c$ is the sample critical temperature. This allows us to fix the peak position in the spectral density as well as the value of $PG_{loss}$ in the pseudogap density of state at 16.1 meV read off the top frame of figure \ref{fig4}. This leaves a single parameter in the characterization of $\tilde{N}(\omega)$. Recently H\"{u}fner et al.\cite{hufner:2008} have provided a summary of known pseudogap values as a function of doping ($p$) for a great variety of systems and from many different measurement techniques. They find that the pseudogap becomes zero only at the top of the superconducting dome and that it grows roughly linearly as doping ($p$) is decreased. We can use their average fit to the data to estimate the pseudogap value in the UD69 sample and obtain $\Delta_{pg} \cong 56$ meV. This fixes our pseudogap density of state model completely and $N_0 = 0.55$. The remaining parameters in $I^2\chi(\omega)$ are then varied, and we get the solid blue curve in the upper frame of Fig. \ref{fig5}. If instead we arbitrarily reduce $\Delta_{pg}$ to 35 meV but change $N_0$ to a value of 0.3 to preserve $PG_{loss}$ at 16.1 meV, the dash-dotted red curve is obtained which shows that the recovered $I^2\chi(\omega)$ of a least square fit to optical data is not very sensitive to the value of $\Delta_{pg}$ used, provided $PG_{loss}$ is left fixed.

While we have presented here only the results of a least square fit with fixed model for $\tilde{N}(\omega)$, we know from our results in the bottom frame of Fig. \ref{fig2} that a maximum entropy inversion with this same fixed $\tilde{N}(\omega)$ would return the same $I^2\chi(\omega)$ as the least square fit procedure did. If, however, we had applied to the optical data a maximum entropy inversion assuming instead that the density of state is flat (no pseudogap structure), we would have obtained the dashed blue curve shown in the upper frame of Fig. \ref{fig5}. As before, the peak in this second function has been pushed upward as compared to the input function. When a pseudogap forms, it depresses the scattering in the energy range below $\omega\leq \Delta_{pg}$ and if this is assigned instead to the effect of the boson spectra, it effectively needs to be reduced in that energy region. Further, for energies above $\Delta_{pg}$ it would need to be increased because of the recovery region in $\tilde{N}(\omega)$ from $\omega=\Delta_{pg}$ to $2\Delta_{pg}$ where the DOS is larger than one. The two effects combine to reduce the spectra weight in $I^2\chi(\omega)$ at low $\omega$, compared with its input value, and to increase it in the region of the peak in the dashed blue curve.

This new finding allows us to reassess the case of optimally doped B-2212 OPT96 inverted by maximum entropy in the work of Hwang et al.\cite{hwang:2007} who assumed a flat density of state model. Returning to the curve given in H\"{u}fner et al.\cite{hufner:2008} we estimate that the pseudogap $\Delta_{pg}$ for this sample has a value of 32 meV. A puzzle noted, but not resolved in Ref. [5], is that the peak position in $I^2\chi(\omega)$ obtained in that work, and shown here as the blue dashed curve in the lower frame of Fig. \ref{fig5}, was 60 meV while neutron scattering gives a smaller value of 45 meV. This is due to the existence of a pseudogap in Bi-2212 OPT96 which was not accounted for in the previous maximum entropy inversion. If we take $\Delta_{pg} = 32$ meV then, reference to the top frame of Fig. \ref{fig4} tells us that we should take $PG_{loss} =$ 8.7 meV to get $\omega_{peak} = 45$ meV which implies $N_0 = 0.59$. This gives the solid blue curve for $I^2\chi(\omega)$. Reducing $\Delta_{pg}$ to 20 meV and keeping $PG_{loss}$ the same, leads to the same inverted $I^2\chi(\omega)$ (dash-dotted red curve) whether one uses a least square fit or maximum entropy.

\section{Summary and conclusions}

We have found that including a pseudogap in the inversion process to obtain an electron-boson spectral density from optical data can have a large influence on the shape of the recovered $I^2\chi(\omega)$. This holds whatever the modality used, be it a maximum entropy technique or a least square fit to a parameterized assumed functional form which represents the spectral density we wish to obtain. Conversely, inversions based on a constant density of electronic states in cases when a pseudogap exists will tend to move a peak associated, for example, with coupling to a spin-1 resonance as measured in polarized inelastic neutron scattering experiments, to higher energies and effectively increase its spectral weight in the electron boson function $I^2\chi(\omega)$. Based on this finding we were led to reexamine the presently available inversion of data in optimally doped Bi-2212 OPT96 for which the optical resonance (a large peak in $I^2\chi(\omega)$) was found at 60 meV which is considerably higher than the neutron resonance in this material found at 45 meV. This discrepancy, so far unresolved, here finds a natural explanation. Optimally doped Bi-2212 already has a significant pseudogap. Taking its value from the compilation provided by H\"{u}fner et al.\cite{hufner:2008} and repeating the inversion, we get a new $I^2\chi(\omega)$ with a large peak at 45 meV in agreement with neutrons. An important intermediate result is our finding that the detailed shape of the electronic density of state $\tilde{N}(\omega)$ including a pseudogap does not impact strongly on the position $\omega_{peak}$ of the resonance in $I^2\chi(\omega)$. What is most important is the number of states removed below $\omega = \Delta_{pg}$ which are assumed to pile up above $\Delta_{pg}$ in a recovery region of order, $\omega \simeq 2\Delta_{pg}$.


\begin{acknowledgments}
JH acknowledges financial suport from the National Research Foundation of Korea (NRFK grant No. 20100008552). JPC was supported by the National Science and Engineering Research Council of Canada (NSERC) and the Canadian Institute for Advanced Research (CIFAR).
\end{acknowledgments}


\bibliographystyle{apsrev4-1}
\bibliography{bib}

\end{document}